\documentclass[twocolumn]{aastex701}

\usepackage{amsmath}
\usepackage{lineno}
\usepackage{grffile}
\usepackage{dcolumn}
\linenumbers
\usepackage{bm}
\usepackage{amssymb}
\usepackage{times}
\usepackage{soul}
\usepackage{natbib}
\usepackage{xcolor}
\usepackage{color}
\usepackage{cancel}
\usepackage{graphicx}

\begin{document}
\title{The Role of Inhomogeneities in the Turbulent Accretion of Black Holes}

\author[orcid=0000-0003-4668-8587,sname='Ficarra']{Giuseppe Ficarra}
\affiliation{Dipartimento di Fisica, Università della Calabria, Arcavacata di Rende (Cosenza), 87036, Italy}
\email[show]{giuseppe.ficarra@unical.it}

\author[orcid=0009-0006-2660-2212,sname='Arcuri']{Michele Arcuri}
\affiliation{Dipartimento di Fisica, Università della Calabria, Arcavacata di Rende (Cosenza), 87036, Italy}
\email{Michelearcuri29@gmail.com}

\author[orcid=0009-0004-9184-5099,sname='Megale']{Rita Megale}
\affiliation{Dipartimento di Fisica, Università della Calabria, Arcavacata di Rende (Cosenza), 87036, Italy}
\email{rita.megale@unical.it}

\author[orcid=0000-0001-8184-2151,sname='Servidio']{Sergio Servidio}
\affiliation{Dipartimento di Fisica, Università della Calabria, Arcavacata di Rende (Cosenza), 87036, Italy}
\email{sergio.servidio@fis.unical.it}

\begin{abstract}
Observations of supermassive black holes by the Event Horizon Telescope reveal significant inhomogeneities, most likely related to density and magnetic field perturbations. To model these features, we conduct high-resolution 2D general-relativistic magnetohydrodynamics (GRMHD) simulations of a Fishbone-Moncrief torus around a Kerr black hole using the Black Hole Accretion Code \texttt{BHAC}. We compare unperturbed accretion with a case featuring plasma density bubbles with pressure balanced magnetic islands of different amplitudes. Power spectrum analysis of accretion time series, performed via the Blackman-Tukey method, shows that the perturbed case exhibits (1) steeper spectral indices compared to the unperturbed case, deviating from the characteristic $1/\omega$ noise spectrum, and (2) increased correlation times, providing evidence for absorption of macro-structures at the event horizon. Spatial auto-correlation analysis of near-horizon turbulence confirms larger energy-containing coherent structures in the perturbed case altering the accretion rate. These results provide new insights for interpreting observations of supermassive black hole environments, where near-horizon turbulence may play a key role in the accretion process.
\end{abstract}


\section{Introduction}
\label{sec:intro}

\begin{figure*}[ht!]
        \centering
        \includegraphics[width=0.99\textwidth]{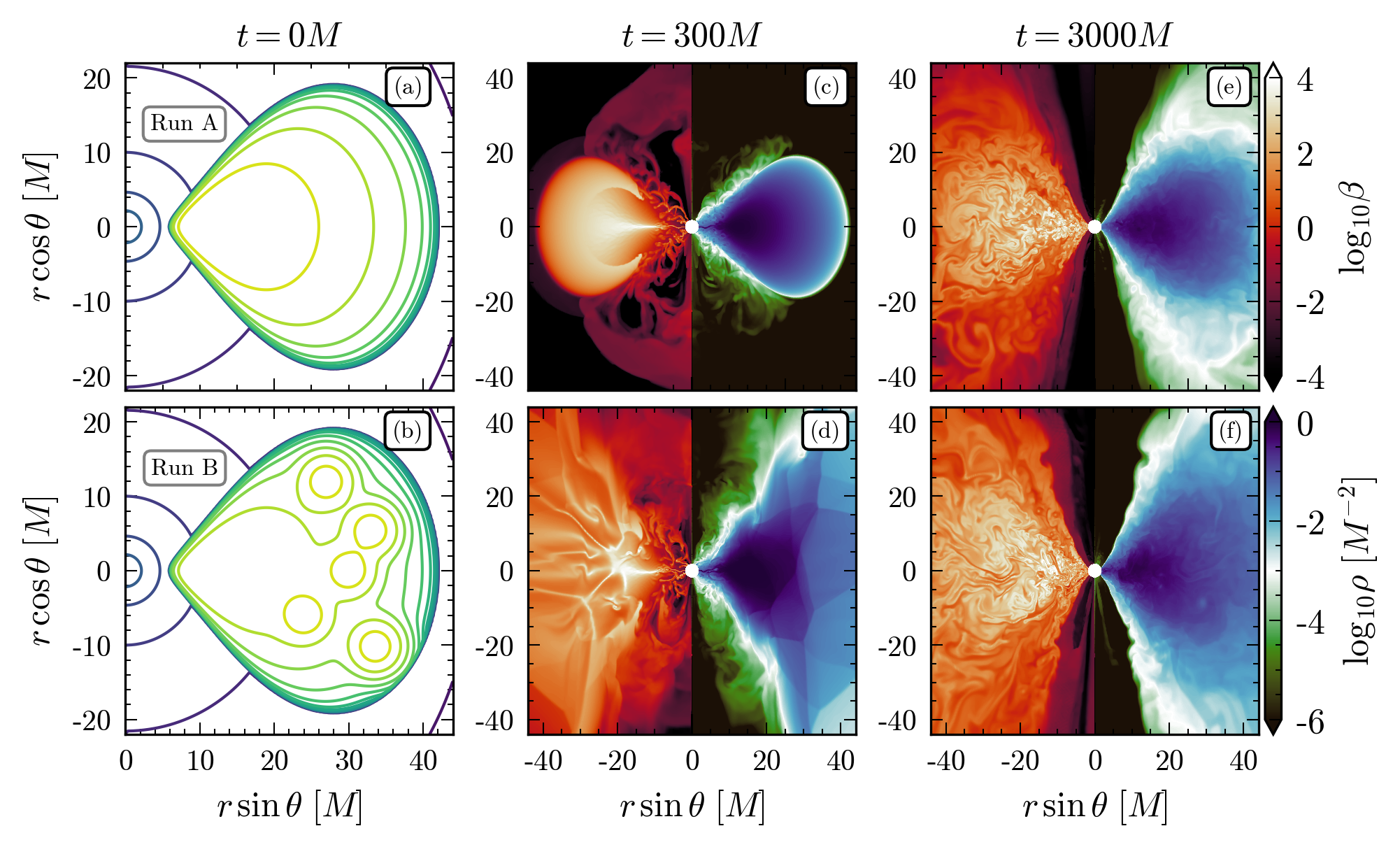}
        \caption{\textit{Left column}: initial density contours for Run A (a) and B (b). \textit{Middle column}: $\log$-snapshots of plasma parameter $\beta$ (left side) and density $\rho$ (right side) at $t=300 \ M$, for Run A (c) and B (d). The white circle represents the black hole event horizon area. \textit{Right column}: same as middle column but at $t=3000 \ M$ for Run A (e) and B (f). In all plots, $r$ and $\theta$ are spheroidal Kerr-Schild coordinates.}
        \label{fig:id_evo_configurations}
\end{figure*}

Supermassive black holes are the largest type of black hole, with masses ranging from hundreds to billions of times that of the Sun. Interstellar gas often gathers around these extremely dense objects, forming rapidly rotating, accretion disks heated to extreme temperatures through friction and compression as matter spirals inward. The ionized nature of the plasma generates strong magnetic fields that extract rotational energy from the black hole to launch powerful collimated jets of particles \citep{BlandfordEA77,BlandfordEA82,Takahashi1990}. In recent years, the Event Horizon Telescope (EHT) captured the first horizon-scale images of the radio sources Sagittarius A* (Sgr A*) ~\citep{EHT_SgrA_PaperI_etal, EHT_SgrA_PaperV_etal} and Messier 87* (M87*) ~\citep{Akiyama2019_L1_etal, Akiyama2019_L5_etal}, supporting the idea that every galaxy might feature a supermassive black hole at its center. These results represent a groundbreaking achievement in high-resolution very-long-baseline interferometry, providing the first direct visual evidence of black holes and enabling studies of the behavior of gravity under extreme conditions.

The observed emission ring surrounding these black holes exhibits pronounced inhomogeneity in brightness, which is primarily attributed to relativistic effects, notably Doppler boosting, as plasma orbiting at near-light speeds emits enhanced radiation when directed toward the observer. Further analysis reveals finer-scale brightness patterns, particularly evident in Sgr A*, indicative of local density and magnetic gradients within the accreting plasma. Self-organizing turbulent processes in the disk might generate such locally dense patterns, which modulate the electromagnetic field geometry and amplify non-uniform emission features across the ring. These structures remain poorly understood and could result from coalescence events, magnetic reconnection \citep{Comisso2018,Parfrey2019,MellahEA22, MeringoloEA23,MeringoloEA25,MeringoloEA26,VosEA23,Vos2024,ImbrognoEA24,ImbrognoEA25} and plasmoid formation \citep{Fermo2010,Uzdensky2010,Huang2012,Loureiro2012,Takamoto2013,Comisso2018}.

In this paper, we analyze the local properties of turbulence in a reduced-dimensionality framework, using a two-dimensional, axisymmetric domain to characterize the statistical properties and topology of coherent structures—such as their typical correlation lengths and the energy-containing scale of magnetic eddies. While this choice allows us to achieve high numerical resolution and robust statistical convergence, it is well known that dimensionality might slightly influence the nature of MHD turbulence. In two dimensions, anti-dynamo theorems constrain magnetic field generation, the magneto-rotational instability saturates differently, and an inverse cascade of magnetic potential can become prominent. Our analysis in this first exploratory study therefore focuses on features that are expected to persist in more realistic three dimensional environments, particularly under the influence of a net mean magnetic field out of the poloidal plane, which tends to render turbulence quasi-two-dimensional in such a plane.

This work investigates how strong, multiple density inhomogeneities with pressure balanced magnetic vortices influence black hole accretion phenomenology and whether small-scale plasma anomalies produce macroscopic effects. We perform general-relativistic magnetohydrodynamic (GRMHD) simulations of accretion disks around Kerr black holes using the open-source Black Hole Accretion Code \texttt{BHAC} \citep{PorthEA2017,OlivaresEA2019}. By comparing simulations with and without initial plasma blobs, we analyze the stochastic properties of the accretion rate and magnetic flux at the event horizon, probing turbulence signatures associated with density and magnetic field inhomogeneities.

\section{Methods and Simulations}
\label{sec:method}

In a strong gravity environment, plasma dynamics are described via ideal GRMHD equations:
\begin{eqnarray}
&&  \nabla_{\mu} ( \rho u^{\mu})=0,\\
&&  \nabla_{\mu} T^{\mu \nu}=0,\\
&&  \nabla_{\mu} {}^{*}F^{\mu \nu}=0,
\end{eqnarray}
where $\rho$ is the rest-mass density, $u^{\mu}$ the fluid four-velocity, $T^{\mu \nu}$ the energy-momentum tensor, and ${}^{*}F^{\mu \nu}$ the dual Faraday tensor. We set geometric units $G=c=1$ and adopt Lorentz-Heaviside units for the definition of the electromagnetic fields. Numerical integration of these equations in arbitrary stationary spacetimes is performed using \texttt{BHAC} \citep{PorthEA2017}. We employ a 2D geometry with Kerr background in logarithmic Kerr-Schild coordinates \citep{MisnerEA73,FontEA98,McKinneyEA04}, enabling extended radial coverage. \texttt{BHAC} implements second-order, high-resolution shock-capturing finite-volume methods, standard practice for localized matter fields. The scheme includes adaptive mesh refinement (AMR) via a block-based strategy and constrained transport \citep{OlivaresEA2019} that maintains the magnetic field's divergence-free condition to machine precision \citep{DelZannaEA2007}.

The accretion disk is modeled as a Fishbone-Moncrief torus \citep{FishboneMoncrief76} with inner radius $r_{\text{in}}=6 \ M$ and density maximum at $r_{\text{max}}=12 \ M$, orbiting a Kerr black hole with dimensionless spin $a_*=0.9375$ and mass $M=1$ (event horizon at $\sim 1.35 \ M$). The magnetic field is initialized via a single poloidal loop vector potential \citep{PorthEA19}, $A_{\phi} = \text{max} \left [ \rho/\rho_{\text{max}} - 0.2, 0 \right ]$, where $\rho_{\max}$ is normalized to unity. An ideal gas equation of state ($\gamma=5/3$) is adopted. The computational grid uses base resolution $(N_r, N_{\theta})=(256, 128)$ with 5 AMR levels, yielding effective resolution $4096\times 2048$ at maximum refinement. In the code workflow, refinement levels are set using Lohner's error estimation scheme \citep{LOHNER1987323}, a well established procedure for codes based on finite volume formulations \citep{MignoneEA2012,ZanottiEA2015}.

\begin{figure*}[t!]
	\centering
	\includegraphics[width=0.99\textwidth]{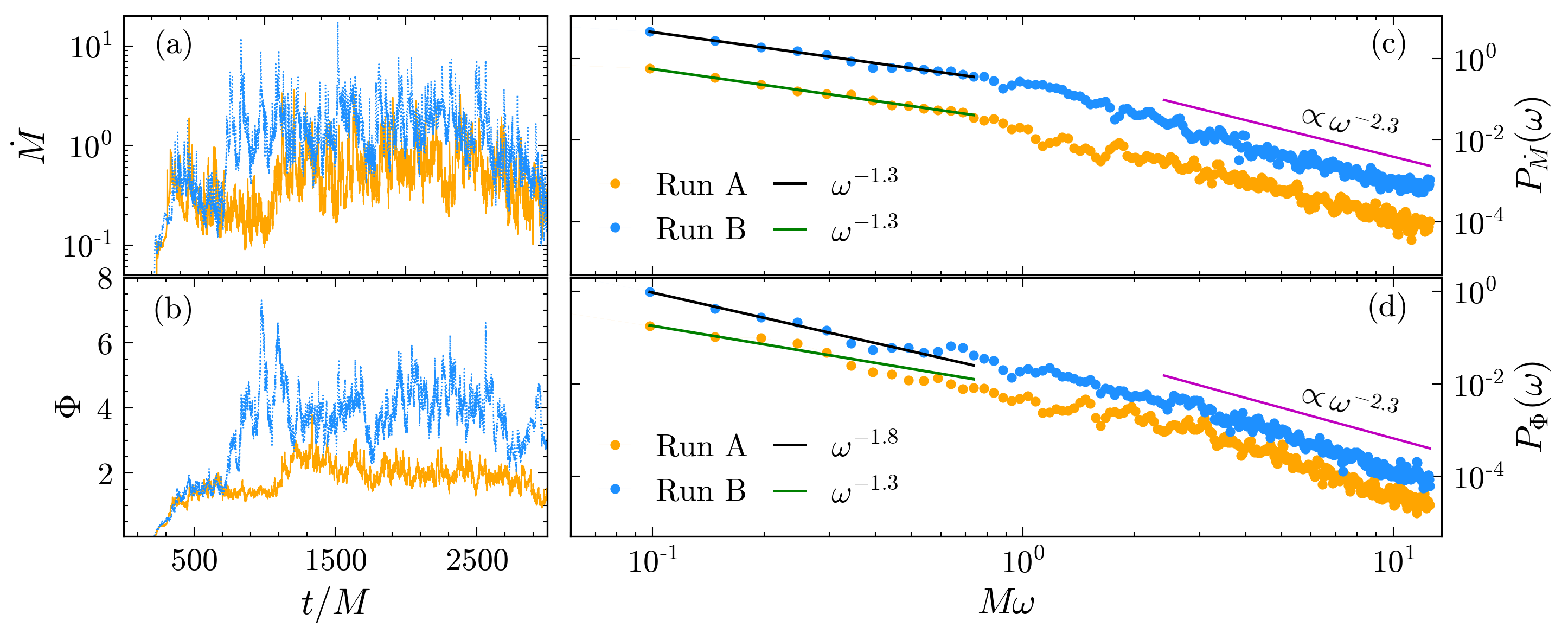}
	\caption{\textit{Left column}: rest-mass accretion rate $\dot{M}$ (a) and accreted magnetic flux $\Phi$ (b) at the black hole horizon. Run A is depicted with solid orange lines, and Run B with dotted blue lines. \textit{Right column}: power spectra of the accretion rate (c) and magnetic flux (d). In both panels, orange and blue dots denote data points from Runs A and B, respectively, while black and green solid lines indicate the corresponding power-law fits at low frequencies. The solid magenta line represents a general tendency common to both configurations at high frequencies.}
	\label{fig:time_spectra}
\end{figure*}

To quantify the role of inhomogeneities in the evolution of the system, we examine two configurations: an unperturbed smooth profile, Run A \citep{PorthEA19}, and Run B, where five plasma bubbles were added to the smooth profile. Specifically, we consider blobs of gaussian shape with fixed width $0.3M$ and amplitude $0.1\rho_{\rm{max}}$ at different locations across the disk, adding them on top of the unperturbed density profile from Run A. Additionally, to introduce more physically meaningful perturbations, we impose pressure balance. Consequently, each density inhomogeneity retains a closed magnetic flux, $A_\phi$, forming a localized structure in the magnetic field, often referred to as a magnetic island or vortex.

The initial density distributions are shown in Fig.~\ref{fig:id_evo_configurations} [(a)-(b)], with contour plots of $\rho$ for each configuration. The middle column [(c)-(d)] displays the plasma $\beta$ parameter ($\beta = 2p/b^2$, where $p$ is pressure and $b=\sqrt{b_\mu b^\mu}$ is the comoving magnetic field strength) and rest-mass density $\rho$ at $t=300 \ M$. The last column [(e)-(f)] show the same quantities at $t=3000 \ M$. Significant differences emerge between the two cases at this intermediate stage and in the final turbulent accretion state, demonstrating persistence of the initial inhomogeneous patterns. The intermediate case with inhomogeneities [panel (d)] exhibits transient compressive activity, characteristic of enhanced magnetosonic modes excited by the initial plasma perturbations. Generally, the system reaches a quasi-steady state in both cases, with patterns typical of MHD turbulence, as illustrated in panels (e)-(f) of Fig.~\ref{fig:id_evo_configurations}. Specifically, the density field displays distinctive blobs and spiral-like features, indicating the presence of vortices within both the disk and wind regions \citep{VosEA23,MeringoloEA23,ImbrognoEA24}.

\begin{figure*}[ht!]
	\centering
	\includegraphics[width=0.99\textwidth]{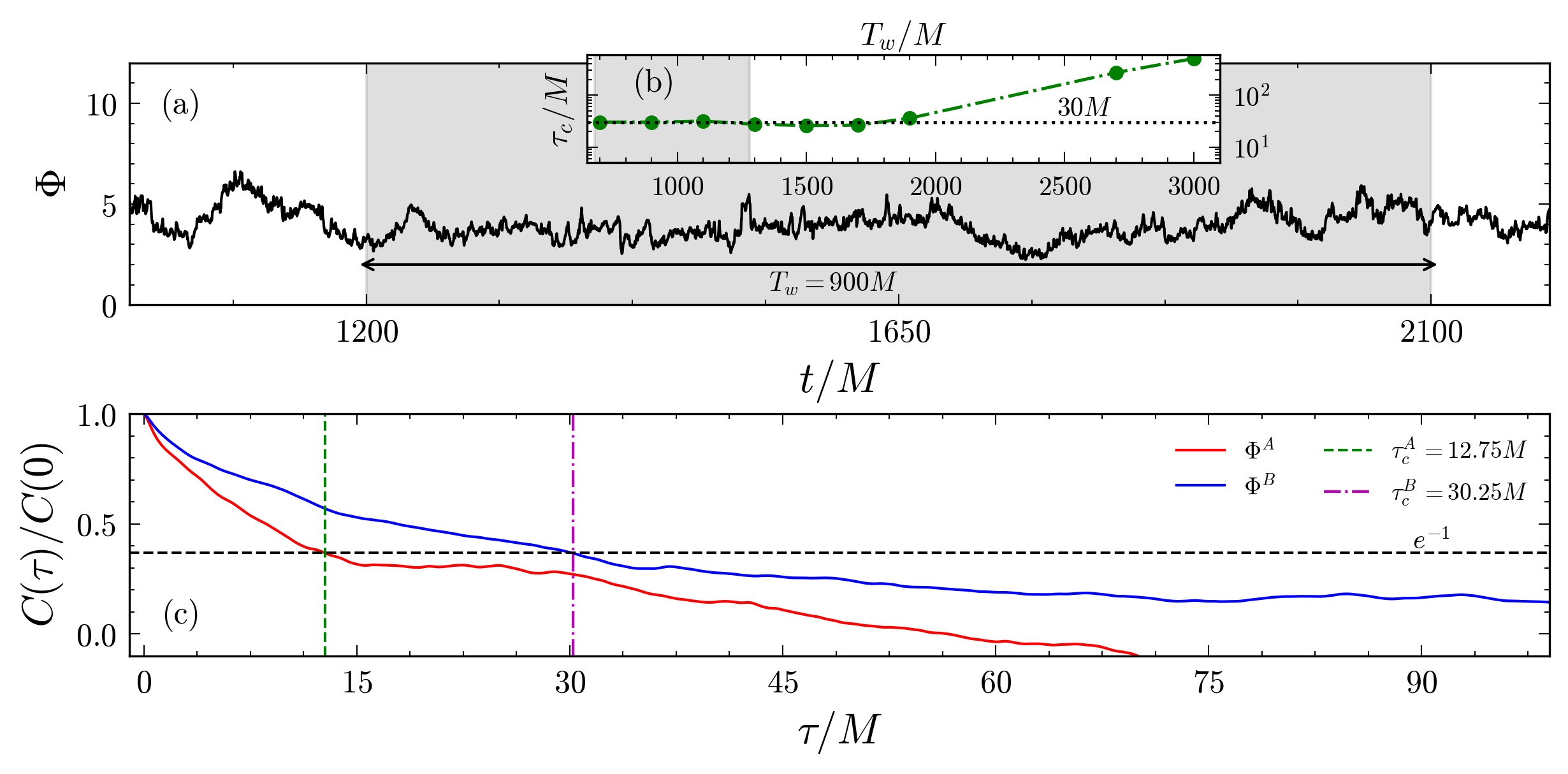}
	\caption{\textit{Top}: Magnetic flux $\Phi$ for Run B (a). The solid black line shows the data and the shaded region indicates a selection window $T_w = 900\,M$. The inset (b) displays a convergence test of the correlation time $\tau_c$ as a function of the selection window length $T_w$. \textit{Bottom}: normalized auto-correlation function $C(\tau)/C(0)$ of the magnetic flux $\Phi$ (c). Solid red and blue lines represent Runs A and B, respectively. Vertical dashed lines show correlation times $\tau_c^{A} = 12.75\,M$ (green) and $\tau_c^{B} = 30.25\,M$ (magenta), while the horizontal dashed black line denotes the $1/e$ threshold.}
	\label{fig:time_correlation}
\end{figure*}

\section{Results}
\label{sec:results}

To compare the two configurations, we first examine global quantities such as the rest-mass accretion rate at the horizon,
\begin{equation}
\dot{M} := \int_{0}^{2\pi}\int_{0}^{\pi} \rho u^r\sqrt{-g} \,d\theta d\phi,
\end{equation}
and accreted magnetic flux
\begin{equation}
\Phi := \frac{1}{2}\int_{0}^{2\pi}\int_{0}^{\pi} |B^r| \sqrt{-g} \,d\theta d\phi.
\end{equation}
These quantities are shown versus time in panels (a)-(b) of Fig.~\ref{fig:time_spectra}. Initial transients persist until $\sim 1200 \ M$ (Run A) and $\sim800 \ M$ (Run B). The inhomogeneous initial conditions yield systematically higher values in both time series - Run A rates are consistently lower than those of Run B, particularly evident in $\Phi$. This indicates enhanced matter inflow in the perturbed case. Furthermore, the perturbed configuration exhibits noticeably more efficient magnetic activity compared to the unperturbed case. The temporal evolution reveals differences not only in magnitude but also in the modal distribution of these rates, as discussed below.

To identify turbulence imprints, we performed a statistical analysis of the fluctuations in these time series and computed their power spectrum via the Blackman-Tukey theorem \citep{BlackmanTukey1958,Matthaeus1982a,Frisch1995,Pecora2023}. The auto-correlation function $C(\tau)$ for a generic field $f(t)$ is defined as
\begin{equation}
C(\tau) = E - \frac{S_2(\tau)}{2}, \label{eq:autocorr}
\end{equation}
where $E$ represents the field variance and $S_2(\tau)=\langle |f(t+\tau)-f(t)|^2\rangle_{T_w}$ denotes the second-order structure function averaged over time window $T_w$. Under the ergodic theorem, this time average equals ensemble averaging when $T_w$ substantially exceeds typical correlation timescales \citep{orszag1974lectures}. 
The power spectrum is then obtained via Fourier transform:
\begin{equation}
P_f(\omega) = \int_{-\infty}^{\infty} C(\tau) \ W_\tau \ e^{-i\omega \tau} \ d\tau,
\end{equation}
with $W_\tau$ being an appropriate window.


Panels (c)-(d) of Fig.~\ref{fig:time_spectra} display power spectra for both the accretion rate ($P_{\dot{M}}$) and magnetic flux ($P_{\Phi}$). At low frequencies ($M\omega\lesssim0.6$), power-law fits $\omega^{\alpha}$ yield $\alpha=-1.3$  for the mass accretion rate in both cases, somehow close to white noise ($\alpha\sim-1$), as observed in stochastic systems \citep{matthaeusEA07,dmitrukEA11}. At high frequencies ($M\omega\gtrsim2$), all profiles converge to a universal $\omega^{-2.3}$ scaling, symptomatic of a process that is probably self-similar \citep{chenEA89}. The magnetic flux spectra show steeper slopes and a net difference among each configuration. At low frequencies, indeed, the systematically steeper spectrum in the inhomogeneous case (Run B) suggests the presence of larger times events advecting across the horizon, consistent with enhanced turbulent injection and coalescence effects \citep{ImbrognoEA24}.

\begin{figure*}[ht!]
	\centering
	\includegraphics[width=.99\textwidth]{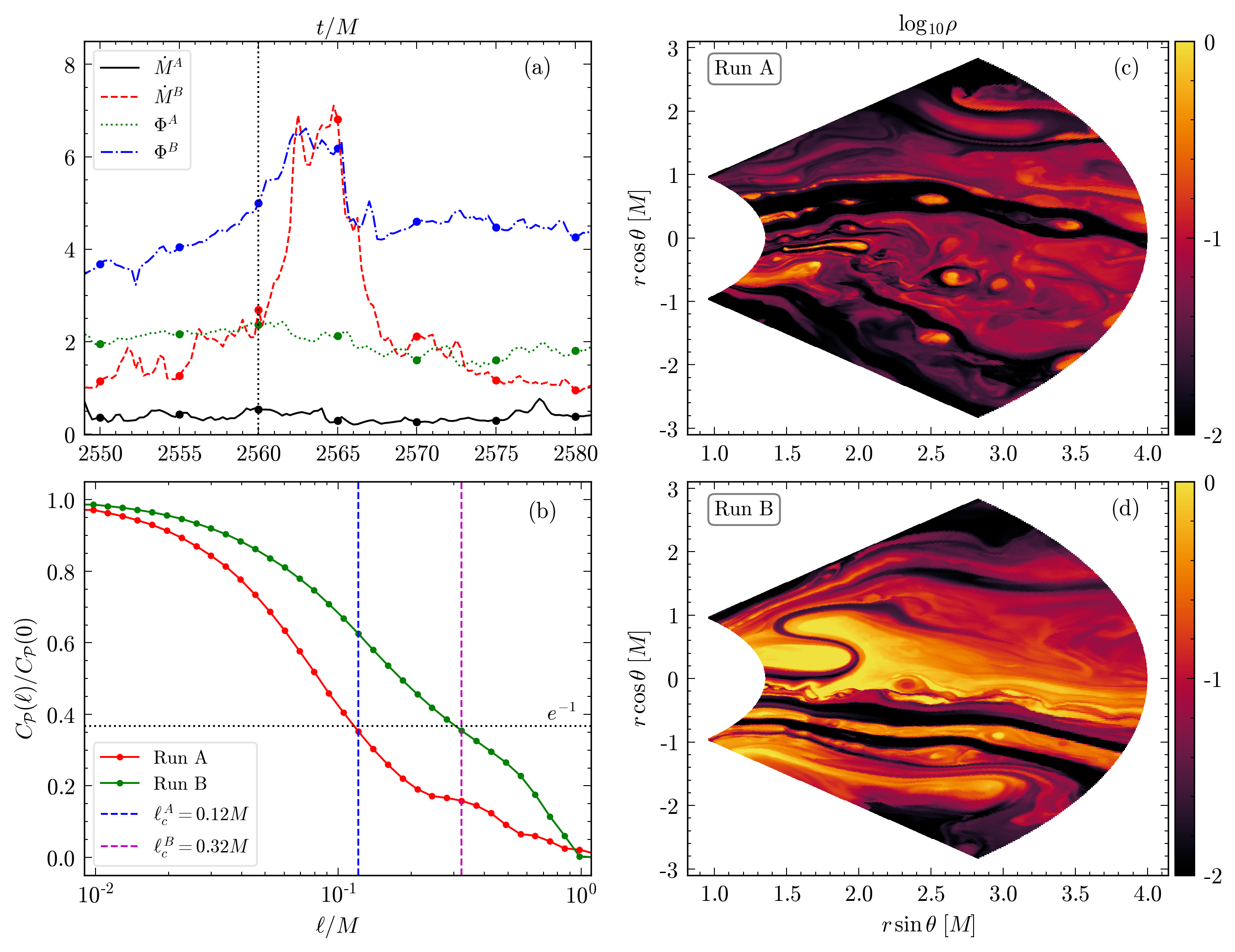}
	\caption{\textit{Left column}: Zoom-in of accretion rate $\dot{M}$ and magnetic flux $\Phi$ around one of the time series peaks (a). The dash-dotted black vertical line shows the time of the spatial analysis $t=2560 \ M$. Normalised auto-correlation function $C(\ell)/C(0)$ (b) of panels (c)-(d). Solid red and green lines depict Runs A and B, respectively. Vertical dashed lines mark $\ell$ values of $0.12\,M$ (blue) and $0.32\,M$ (magenta), while the horizontal dotted black line represents the $1/e$ threshold. \textit{Right column}: Near-horizon density snapshots for Run A (c) and B (d) at $t=2560\,M$. }
	\label{fig:space_correlation}
\end{figure*}

The breaking frequency in Fig.~\ref{fig:time_spectra} [panels (c)-(d)] corresponds to a characteristic timescale derived from the auto-correlation function $C(\tau)$. This timescale estimates the interval over which the system remains strongly correlated, representing an energy-containing timescale. Following standard practice \citep{Frisch1995,dmitrukEA11}, we define the correlation time as the $e$-folding time of the normalized auto-correlation function $C(\tau)/C(0)$ (with $C(0) \equiv E$). However, for systems exhibiting $1/\omega$ noise \citep{ServidioEA08S}, such measurement requires careful consideration of $T_w$, that needs to be sufficiently large to capture correlations but small enough to avoid unsteady effects.

The initial transient trends visible in Fig.~\ref{fig:time_spectra} [panels (a)-(b)] must be excluded from this analysis, since the system is not yet in a quasi-steady state. To address this convergence issue inherent to stochastic signals, we compute $C(\tau, T_w)$ over multiple windows $T_w$ centered at a specific time $T_0$, systematically varying the window size. As an example of the procedure, we consider the time series of the magnetic flux for Run B [panel (a) of Fig.~\ref{fig:time_correlation}] and select $T_0 = 1650 \,M$. Subsequently, we perform a convergence study of the correlation time $\tau_c$ as a function of the selection window length $T_w$ [panel (b) of Fig.~\ref{fig:time_correlation}] which demonstrates that statistics diverge at large $T_w$ due to unsteadiness, while converging properly when $T_w$ approaches $\tau_c$. Finally, we select $T_w=900\,M$, well inside the convergence regime. We apply the same procedure also to the accretion rate $\dot{M}$.  

\begin{table}[htpb!]
	\centering
    \begin{footnotesize}
	\begin{tabular}{| c | c c c c c |}
	\hline
		Run & $\alpha$ & $\beta$ & $\tau_c^{\Phi}$ $[M]$& $\tau_c^{\dot{M}}$ $[M]$\rule{0pt}{2.5ex} & $\ell_c$ $[M]$ \\ [0.5ex]
		\hline\hline
        A (unpert.) & -1.3 $\pm$ 0.1 & -2.2 $\pm$ 0.1 & 12.75 & 7.2 & 0.12 \\ 
		B (pert.) & -1.8 $\pm$ 0.1 & -2.3 $\pm$ 0.1 & 30.25 & 8.5 & 0.32 \\ 
		\hline
	\end{tabular}
    \end{footnotesize}
	\caption{Summary of the analysis results for the two configurations: the unperturbed Run A and the perturbed Run B. The parameters $\alpha$ and $\beta$ represent the spectral slopes of the magnetic flux $\Phi$ at low and high frequencies, respectively. $\tau_c^{\Phi}$ denotes the correlation times of the magnetic flux, while $\tau_c^{\dot{M}}$ the one for the accretion rate. $\ell_c$ is the spatial correlation length of the density $\rho$ computed at $t=2560M$.}
	\label{tab:analysis_results}
\end{table}

Once the appropriate $T_w$ is selected, we measure the correlation times $\tau_c^{{\Phi},\dot{M}}$. The accretion rate $\dot{M}$ exhibits the shortest correlation times, with $\tau_c^{\dot{M}}\sim 7-8 \ M$, as reported in Tab.~\ref{tab:analysis_results}. Magnetic flux timescales are significantly larger, showing a pronounced difference between Runs A and B, with the perturbed case showing consistently longer correlation times, a key finding of this analysis. This peculiar feature directly follows from the fact that $\Phi$ is directly connected to the magnetic field large-scale structures extending from the event horizon to the inner region of the disk, implying longer correlation times. On the other hand, $\dot{M}$ is mostly defined by the small-scale density structures that flow through the event horizon.

It is natural to ask what determines these characteristic, finite timescales, and why the spectral behavior of spatially-averaged quantities differs between Runs A and B. Figure~\ref{fig:time_spectra} reveals more frequent peaks and extreme events in Run B, suggesting these might correspond to different spatial scales of structures crossing the event horizon. During the interval $2550 \ M \leq t \leq 2580 \ M$ [Fig.~\ref{fig:space_correlation}, panel (a)], Run B exhibits prominent growth in both $\dot{M}$ and $\Phi$ peaks, while Run A maintains comparatively random and flat accretion profiles - a characteristic difference evident throughout the time series.

To inspect this behavior, we analyze the corresponding density patterns near the horizon within $1.36 \ M \leq r \leq 4 \ M$ and $\pi/4 \leq \theta \leq 3\pi/4$ at $t=2560 \ M$ (prior to Run B's peak), as shown in panels (c)-(d) of Fig.~\ref{fig:space_correlation}. Both configurations exhibit turbulent patterns (plasmoids, vortices and current sheets) consistent with Fig.~\ref{fig:id_evo_configurations} [panels (e)-(f)], but Run B develops significantly larger blob-like structures, with enhanced ingestion of both matter and magnetic field. This suggests that initial inhomogeneities seed larger coherent structures through turbulent coalescence, enhancing time variability in horizon fluxes. The extreme peaks in the inhomogeneous and perturbed Run B correspond to ``extreme eating events'', where the black hole accretes such macroscopic structures.

To quantify structure sizes and compare smooth versus perturbed configurations, we calculate the spatial autocorrelation function $C_{\mathcal{P}}(\ell)$ of the density field $\rho$ \citep{Frisch1995}, evaluated along all the curves that connect points of the domain separated by a proper length, as described in \citet{MegaleEA25}. Specifically, the invariant distance between two representative points A and B on a constant-time hypersurface is given by 
\begin{equation}
    \ell := \int_{x_A^k}^{x_B^k} \sqrt{\gamma_{ij} dx^i dx^j}\,,
     \label{eq:ell}
\end{equation}
where $\gamma_{ij}$ is the spatial three-metric and $x_{A,B}^k$ are the spatial coordinates of the chosen points. 
This definition describes a geometric property of a given slice that is invariant with respect to a choice of the spatial coordinates parametrization.
We define a proper spatial autocorrelation function as
\begin{equation}
  C_{\mathcal{P}}(\ell) := C_{\mathcal{P}}(0) - \frac{1}{2}\left \langle \frac{1}{\mathcal{V}} \int
  \Delta\rho^2 \ \alpha \ \sqrt{\gamma(x_A^k)}
  \ d^3 x_A \right \rangle_{\Omega}\,,
  \label{eqn:S2gamma+geo}
\end{equation}
where $\Delta\rho = \left| \rho(x_B^k) - \rho(x_A^k) \right|$ and $C_{\mathcal{P}}(0)$ is a proper-volume average of the energy in the fluctuations of
the field or, equivalently, its variance,
\begin{equation}
  C_{\mathcal{P}}(0) := \frac{1}{\mathcal{V}} \int \rho^2 \ \alpha \ \sqrt{\gamma} \ d^3 x\,,
  \label{eq:variance}
\end{equation}
and $\mathcal{V} = \int \alpha \ \sqrt{\gamma} \ d^3 x$ is the proper volume.
Additionally, $\alpha$ is the lapse function, $\langle\dots\rangle_{\Omega}$ denotes an average over the solid angle and $\gamma$
is the determinant of the spatial metric $\gamma_{ij}$, thus accounting
for local variations of the volume in response to the background
curvature.

This statistical measure characterizes the spatial energy distribution of fluctuations in homogeneous turbulence in curved spacetime, generalizing a standard approach in flat spacetime for both fluid and plasma turbulence studies. We evaluate $C_{\mathcal{P}}(\ell)$ in the near-horizon region [Fig. 4, panels (c)-(d)], to identify energy-containing structures. The normalized $C_{\mathcal{P}}(\ell)/C_{\mathcal{P}}(0)$[Fig.~\ref{fig:space_correlation}, panel (b)] yields correlation lengths $\ell_c = 0.12 \ M$ (Run A) and $\ell_c=0.32 \ M$ (Run B), defined at the $e^{-1}$ decay point. This refined analysis, summarized in Tab. \ref{tab:analysis_results}, demonstrates how the correlation length depends strongly on the accretion regime, with larger structures emerging in the perturbed case. This size enhancement reflects coalescence processes in magnetized turbulence, where magnetic helicity drives flux tube growth \citep{ImbrognoEA24} - an inverse transfer that is amplified in strongly perturbed disks.

\section{Discussion}
\label{sec:discussion}
We performed GRMHD simulations of accretion disks with plasma perturbations, revealing that initial density inhomogeneities with pressure balanced magnetic islands influence the system evolution. Specifically, randomly sized and amplitude plasma bubbles were introduced to study plasma dynamics under strong gravity and magnetic fields near compact objects. We analyzed the stochastic properties of both the accretion rate and magnetic flux at the black hole horizon to characterize time variability. Power spectrum analysis shows the magnetic flux accretion depends strongly on initial configuration, with substantially different slopes at low accretion frequencies $\omega$. 

Since the low-frequency behavior depends on the initial configuration, we performed a correlation time study of these signals to quantify the typical temporal scale of large accretions. The characteristic correlation time of the magnetic flux is about three times larger in the perturbed case compared to the unperturbed regime. The mass accretion rates also have bigger correlation times in the perturbed case. Such results suggest that the density perturbations initiate a different regime of the turbulent cascade.

To gain more insight into the amplification of accretion times, we linked the time variability of these signals to large-scale turbulence properties in the near-horizon disk. Through proper autocorrelation function analysis in curved spacetime using a novel technique, introduced recently in \citet{MegaleEA25}, we measured the spatial correlation length - a proxy for energy-containing eddies. This innovative analysis is well suited to analyze the selected near-horizon region, where the spacetime curvature is the strongest, and the modified autocorrelation function accounts for this by incorporating the spatial metric on the constant-time hypersurface. Such diagnostic tool shows that initial inhomogeneities lead to larger-scale structures, consistent with enhanced energy and magnetic helicity turbulent energy cascade. Globally, this amplified magnetic and fluid activity produces larger variability in both the accretion rate and magnetic flux, with macro-structures penetrating the horizon over longer times, leading to increased correlation times. These findings suggest that small-scale turbulence has a crucial effect also on large-scale and low frequency observables.

Our study introduces a controlled prescription of density perturbations, with associated magnetic field islands, into a classical black hole accretion model. While the specific size and distribution of these blobs are chosen for numerical tractability, they are motivated by several astrophysical scenarios. Observations of sources like Sgr A* and M87 reveal inhomogeneous, clumpy structures in their accretion flows, with emission (and hence density) fluctuations on scales comparable to a significant fraction of the disk diameter \citep{Akiyama2019_L1_etal,EHT_SgrA_PaperI_etal}. Such overdensities could originate from various processes, including: the tidal disruption and accretion of stars or stellar debris \citep{StoneEA20}; localized inhomogeneities advected from the galactic environment or neighboring interstellar clouds \citep{KormendyEA13}; or the formation and injection of magnetized plasmoids via magnetic reconnection events within the disk itself \citep{Beloborodov17,SironiEA19}. These processes can naturally produce density contrasts and coherent structures of the order implemented in our simulations. Our simplified model thus serves as a foundational investigation into how discrete, mesoscopic perturbations influence global accretion dynamics, turbulence, and horizon-scale variability.

The introduction of plasma bubbles modifies the initial profile of the magnetic field, similarly to the case of multi-field loop configurations, which have been extensively analyzed in the context of accretion physics, both in the force-free limit \citep{ParfreyEA15, YuanEA19a, YuanEA19b, MahlmannEA20} and in fully relativistic MHD \citep{BarkovEA11, ChristieEA19,NathanailEA20,NathanailEA22}. The main effect of the magnetic field geometry is observed in the presence of loops with alternating or multiple polarities, leading to large magnetic activity and generation of extremely powerful jets. In our case the magnetic initial seed presents a unique polarity, typical of SANE accretion models, that are featured by magnetized funnels with much smaller magnetic flux and jet power compared to magnetically arrested disk (MAD) models or multi-polarity cases. Additionally, the normalized magnetic flux at the horizon $\phi = \Phi/\sqrt{\dot{M}}$ never exceeds a value of $\approx 8$ for Run B, which is smaller than the typical value $\approx 15$ for a MAD accretion state \citep{TchekhovskoyEA11}. Consequently, we expect differences observed between Runs A and B mostly arise from the amplitude of the density perturbations themselves, rather than the specific topology of the magnetic initial seed.

The reduced dimensionality of our model can influence certain aspects of turbulent dynamics—such as the elongation of structures along the mean field, their disruption via 3D vorticity stretching and 3D magnetic reconnection. However, our 2.5D setup captures the spectral energy transfer in the plane perpendicular to the mean magnetic field, especially in the disk and wind regions, providing a well-resolved and spectrally convergent description of turbulence in these zones. In the presence of a mean field, the analogies between 2D and magnetized 3D are well documented in relativistic turbulence \citep{ComissoEA19}. We emphasize that the large-scale, coherent flux ropes observed in our simulations have direct analogues in 3D simulations of accretion flows, where similar helical structures emerge from turbulent dynamics under the constraint of magnetic helicity conservation, as observed in 3D simulations of general relativistic particle in cell plasmas \citep{MellahEA23} and 3D GRMHD simulations of magnetized accretion flows \citep{NathanailEA22}. Such structures have also been widely documented in heliospheric plasmas, supporting the physical relevance of the phenomena reported here \citep{PecoraEA19}. Finally, the covariant formalism we employ for computing structure functions and power spectra is inherently three-dimensional, as it operates with proper lengths and volumes. This framework can be directly applied to future high-resolution 3D simulations, enabling a direct comparison of 2D versus 3D turbulent statistics in strongly curved spacetimes.

\begin{acknowledgments}
The authors are grateful to the anonymous Reviewer for a careful reading of the manuscript and for providing insightful comments and valuable suggestions, which have significantly improved the presentation of this﻿ work. G. F. gratefully acknowledges the support of University of Calabria through a research fellowship funded by DR 1688/2023. R. M. and S.S. acknowledge the support provided by ``ICSC - Centro Nazionale di Ricerca in High Performance Computing, Big Data and Quantum Computing - and hosting entity, funded by European Union-NextGenerationEU''. Authors acknowledge computational resources for simulations and analysis from the Alarico HPPC Computing Facility at the University of Calabria.
\end{acknowledgments}

\bibliography{BIBLIO_GR}
\bibliographystyle{aasjournalv7}



\end{document}